\documentclass[preprint,authoryear,12pt]{elsarticle}

\usepackage{graphicx}
\usepackage{epsfig}

\usepackage{amssymb}

\usepackage[ps2pdf,%
a4paper=true,%
breaklinks=true,%
colorlinks=true,%
pdfauthor={First Author et al.},%
pdftitle={Template for manuscripts in Advances in Space Research}%
]{hyperref}

\journal{Advances in Space Research}

\begin{document}

\begin{frontmatter}



\title{Mapping lunar surface chemistry: new prospects with the Chandrayaan-2 Large Area Soft x-ray Spectrometer (CLASS)}


\author[label1]{S. Narendranath\corref{cor}}

\cortext[cor]{Corresponding author}
\ead{kcshyama@isac.gov.in}


\author[label1,label3]{P.S. Athiray}
\author[label1]{P. Sreekumar}
\author[label1]{V. Radhakrishna}
\author[label1]{A. Tyagi}
\author[label2]{B.J. Kellett}
\author[label1,label4]{and the CLASS team}
\address[label1]{Space Astronomy Group, ISITE Campus, ISRO, Marathalli Ring Road, Bangalore 560037, India.}

\address[label2]{STFC, Rutherford Appleton Laboratory, Chilton, UK.}
\address[label3]{Department of Physics, University of Calicut, Malappuram, Kerala}
\address[label4]{e2V centre for electronic imaging, Planetary and Space Science Research Institute, Open University, Milton Keynes, UK. }
\begin{abstract}

Surface chemistry of airless bodies in the solar system can be derived from remote x-ray spectral measurements from an orbiting spacecraft. X-rays from planetary surfaces are excited primarily by solar x-rays. Several experiments in the past have used this technique of x-ray fluorescence for deriving abundances of the major rock forming elements. The Chandrayaan-2 orbiter carries an x-ray fluorescence experiment named CLASS that is designed based on results from its predecessor C1XS flown on Chandrayaan-1. We discuss the new aspects of lunar science that can be potentially achieved with CLASS.


\end{abstract}


\end{frontmatter}

\parindent=0.5 cm

\section{Introduction}
 
Characteristic x-rays excited by solar x-rays or an active source (when measurements are done in-situ) provide signatures of elements in the upper few hundred microns of the regolith. There has been several experiments in the past (Figure 1) from which surface compositions were derived with this technique. One of the major constraints for orbital experiments is the signal levels that are entirely dependent on solar activity. Chandrayaan-1 X-ray Spectrometer (C1XS) \citep{Grande09} though it performed well could detect XRF signals only during a few flares due to the low solar activity in 2009. 

\par In this paper we discuss the potential science capabilities of the Chandrayaan-2 Large Area Soft x-ray Spectrometer (CLASS) \citep{RK11} to be flown on the Indian lunar mission Chandrayaan-2. 
\begin{figure}
\includegraphics[height=5cm,width=8cm]{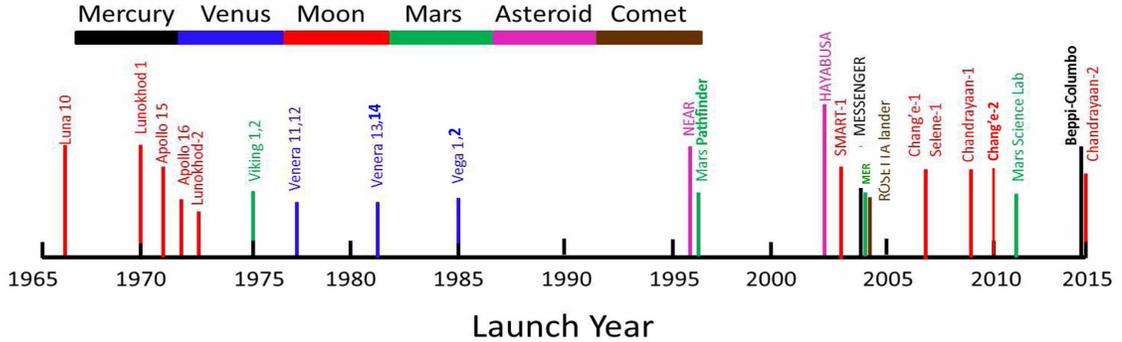}
\caption{Historical time line of XRF based experiments for solar system studies (includes experiments with an active source on landers/rovers. For long duration missions, the launch dates are indicated.}
\end{figure}

\subsection{Instrument}
CLASS will make use of sixteen large area swept charge devices  \citep{L2001} termed 'CCD-236' \citep{Br} manufactured by e2V Limited, UK. Each detector has an area of 4 cm$^2$ and together provide an active geometrical area of 64 cm$^2$.  CLASS will operate in the 0.8- 15 keV energy range. An energy resolution less than 250 eV at 1.48 keV is targeted at an operating temperature of -20 $^{\circ}$C. The changes in CLASS with respect to C1XS is listed in table 1. 
\par  Copper collimator coated with gold will be used to restrict the field of view to $\sim$ 14$^{\circ}$ (full opening angle) which would define a ground pixel of 25 km x 25 km (FWHM) on the lunar surface from the 200 km orbit of Chandrayaan-2. The detectors are arranged in four quadrants as shown in figure 2. The housing is aluminum with a thick door to protect the devices from radiation damage en-route to the Moon. Four Fe-55 radioisotopes with Ti foils are embedded on the door so that each quadrant is illuminated by one source for on-board calibration (providing lines at energies 4.5keV, 4.9 keV, 5.9 keV and 6.4 keV) when the door is closed.


\begin{table}

\caption{Comparison of C1XS and CLASS}
\begin{tabular}{lll}
\hline
{\bf Parameter}&{\bf C1XS}&{\bf CLASS}\\
\hline
Geometric area & 24 cm$^2$ & 64 cm$^2$\\ 
                & 1 cm$^2$ x 24 & 4 cm$^2$ x 16 \\ \hline
X-ray detector& CCD-54&CCD-236 \\ \hline
Foot print&50 km x 50 km & 25 km x 25 km \\ 
(at 200 km)&&\\ \hline
Spectral res&143 eV&$<$ 200 eV \\ 
at 5.9 keV;-20 $^{\circ}$C && \\ \hline
Energy range&1-20 keV&0.9-15 keV \\ \hline

\end{tabular}
\label{table1}

\end{table}

\begin{figure}
\begin{center}
\includegraphics[height=10cm,width=12cm]{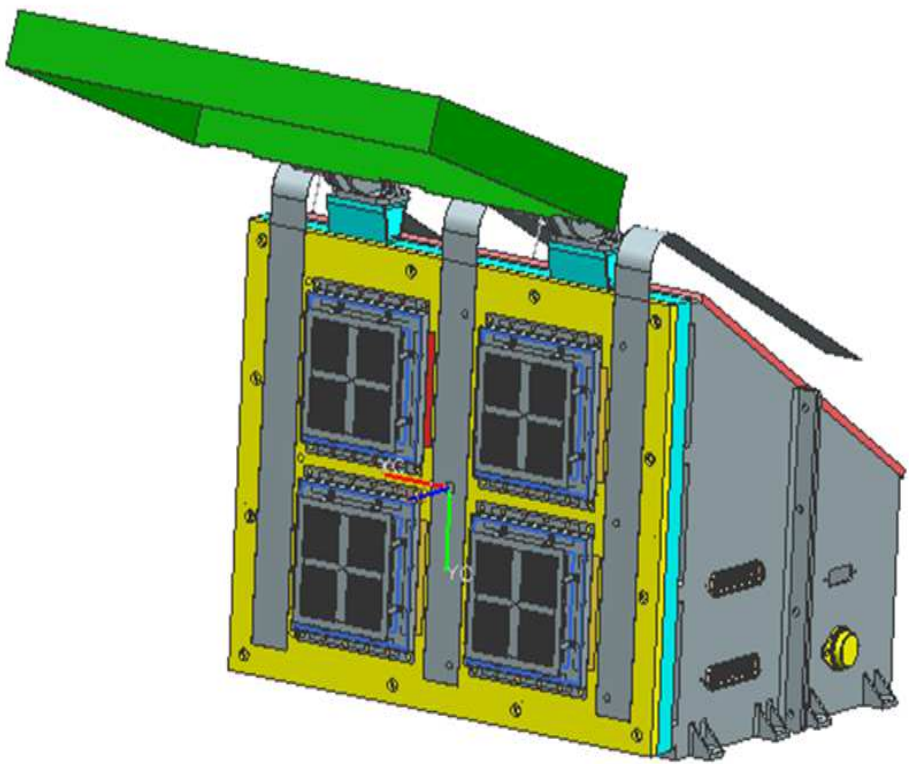}
\end{center}
\caption{CLASS instrument. Each quadrant contains four detectors}
\end{figure}

\section{Lunar compositional diversities with CLASS}

Recent missions such as Kaguya, Chandrayaan-1 and Lunar Reconnaissance Orbiter (LRO) has provided a wealth of data on the surface composition of the Moon. New rock types containing Mg-spinel \citep{pieters2011} and pure crystalline anorthosites (PAN)  \citep{Ohtake2009} are observed. Moon Mineralogy Mapper (M3) on Chandrayaan-1 indicates a latitude-dependent abundance of hydrated phases \citep{pieters2009} hitherto unknown. These point to the fact that the lunar surface composition is far from understood. While it is generally believed that the primary crust of the Moon formed when plagioclase crystallized and floated towards the surface during the late stages of the lunar magma ocean, there are several models with ample uncertainties to describe the structure and evolution of the magma.  Global elemental composition along with mineralogy provides the key for testing such models.

\par Mineralogy has been mainly studied from visible to near infrared reflectance spectroscopy from orbit while elemental composition is studied through x-ray or gamma ray measurements. The major science objectives of such experiments largely remain the same (for example, see \citep{crawford2009}). Here we highlight only the new science aspects that can be done with elemental maps from CLASS.

The most common mineral in highlands is feldspar belonging to plagioclase series which are solid solutions between albite (NaAlSi$_{3}$O$_8$) and anorthite (CaAl$_2$Si$_2$O$_8$). The plagioclase composition can be characterized by measuring the ratio Ca/(Ca+Na+K) (expressed in moles) known as the An number (AN$\#$). K is found only as a trace element in lunar feldspars and hence the ratio Ca/(Ca+Na) (both major elements) would be a good indicator of AN$\#$. Plagioclase in the returned lunar samples has an anorthite content rather narrow in range from which the average highlands are estimated to be An$_{95}$.  Plagioclase grains with AN$\#$ as low  An$_{70}$ \citep{Wz2006} have been found in lunar samples though rare.  
\par Global mineralogy has been understood largely through remote sensing measurements in UV/VIS and IR. In these methods, the anorthite content cannot be estimated unless the plagioclase is of the purest form as was observed by the spectral profiler on Kaguya \citep{Ohtake2009, Yamamoto2012}.  This form of pure crystalline plagioclase is identified by the 1.25 $\mu$m broad absorption feature (which arises from the minor amounts of Fe$^{2+}$ in the crystal structure) whose characteristics (like band center , depth) can be used to derive the AN$\#$. Otherwise, plagioclase on the Moon is identified by a featureless near-IR spectrum which does not reveal the An content.
While it is possible to measure major elements with gamma ray spectroscopy, detection of Na is not possible since there are no gamma rays lines for Na. X-ray measurements can provide Na abundances necessary for studying plagioclase compositions.

\par X-ray fluorescence (XRF) observations measure the abundance irrespective of the mineral structure. XRF spectral analysis can uniquely identify and quantify elemental signatures from all commonly occurring elements (depending on the sensitivity of the experiment).  Remote sensing XRF experiments are limited by the detector window absorption at lower energies and effective collecting area of the detector and hence can measure elements from Na to Fe if they are present at above $\sim$ 1 wt$\%$. X-ray fluorescence data from C1XS on Chandrayaan-1 indicate the presence of 1-2 wt$\%$ of Na in some of the sampled highlands \citep{Naren2011}. It was also found that Al is as high as 18 wt$\%$ while Ca is in the range of 6-9 wt$\%$. Observations by the Diviner radiometer on LRO, which has three IR channels near 8 $\mu$m to measure the Christiansen feature (CF), has revealed regions where the band center shifts beyond their estimated range for plagioclase, suggestive of the possible existence of higher amounts of sodic plagioclase \citep{Green2010}. Global distribution of pure anorthositic material \citep{Yamamoto2012} by Kaguya reveal regions of high Al similar to C1XS, extending to kms but there is no direct quantitative measure (the Al abundance quoted are model dependent).
\par Given the new set of data suggesting variations in the plagioclase composition globally,  one of the major objectives of CLASS is to measure Na in the lunar surface. Simultaneous measurements of Na, Ca and Al would aid in revealing the diversity in plagioclase compositions. Figure 3 shows simulated lunar XRF spectra for plagioclase compositions ranging from An$_{60}$ to An$_{100}$ with CLASS during a C3.0 class flare from a 25 km x 25 km ground pixel.
\par We have used the solar spectrum measured at the peak of the C3.0 class solar flare with X-ray Solar Monitor (XSM) on Chandrayaan-1 for this simulation. The XSM spectrum was modeled with a single temperature plasma at 1.07 keV , emission measure of 0.18 x 10$^{49}$ per cm$^{3}$ in \citep{Nrrev}. The lunar spectra for the compositions of interest is simulated with the XSM spectrum as input to a code 'x2abundance' \citep{Ath2013} which we have used for C1XS spectral modeling. The scattered solar x-ray spectrum contains several lines especially in the 1-2 keV energy range which has to be carefully modeled \citep{Wiederb12}. We have calculated the scattered solar spectrum and this is also a component of the simulated spectra along with the typical background in the lunar orbit as observed in C1XS detectors. In the lower panel of figure 3, the flux ratio  Ca/(Ca+Na) derived from fits to the simulated spectra is plotted against the An number showing a good correlation.
 \par With a careful calibration at 1 keV  we hope to determine the abundance of Na along with that of other elements with CLASS data so as to derive a good measure of  AN$\#$ distribution. Further correlation studies with IR spectra can reveal consistency with regions identified as pure crystalline anorthosite.

\begin{figure}
\begin{center}
\includegraphics[height=13cm,width=9cm,angle=-90]{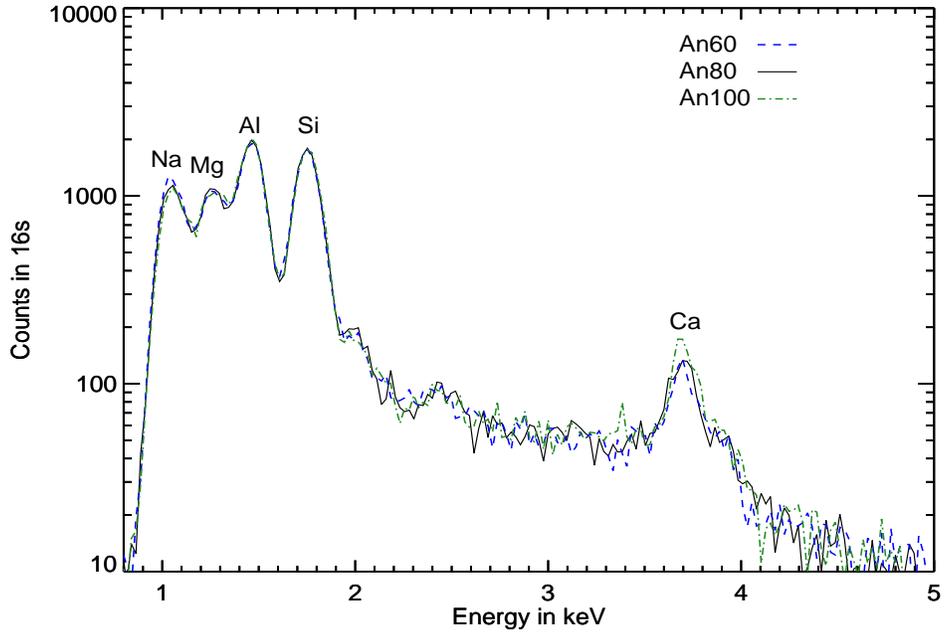}
\includegraphics[height=13cm,width=9cm,angle=-90]{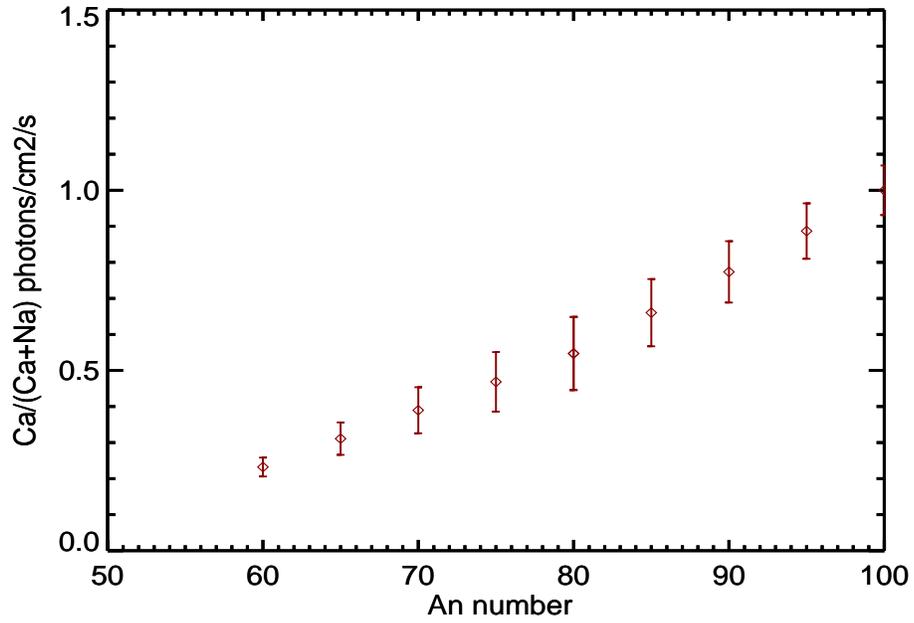}

\end{center}
\caption{Upper panel shows three of the simulated spectra in CLASS with different An numbers for a solar flare of class C3. Lower panel shows the ratio of flux in Ca and Na lines from the simulated spectra and its correlation to the An number}
\end{figure}

\section{Reducing uncertainties}
\subsection{Detector background arising from particles in the lunar orbit}
 
\begin{figure}
\begin{center}
\includegraphics[height=13cm,width=9cm,angle=-90]{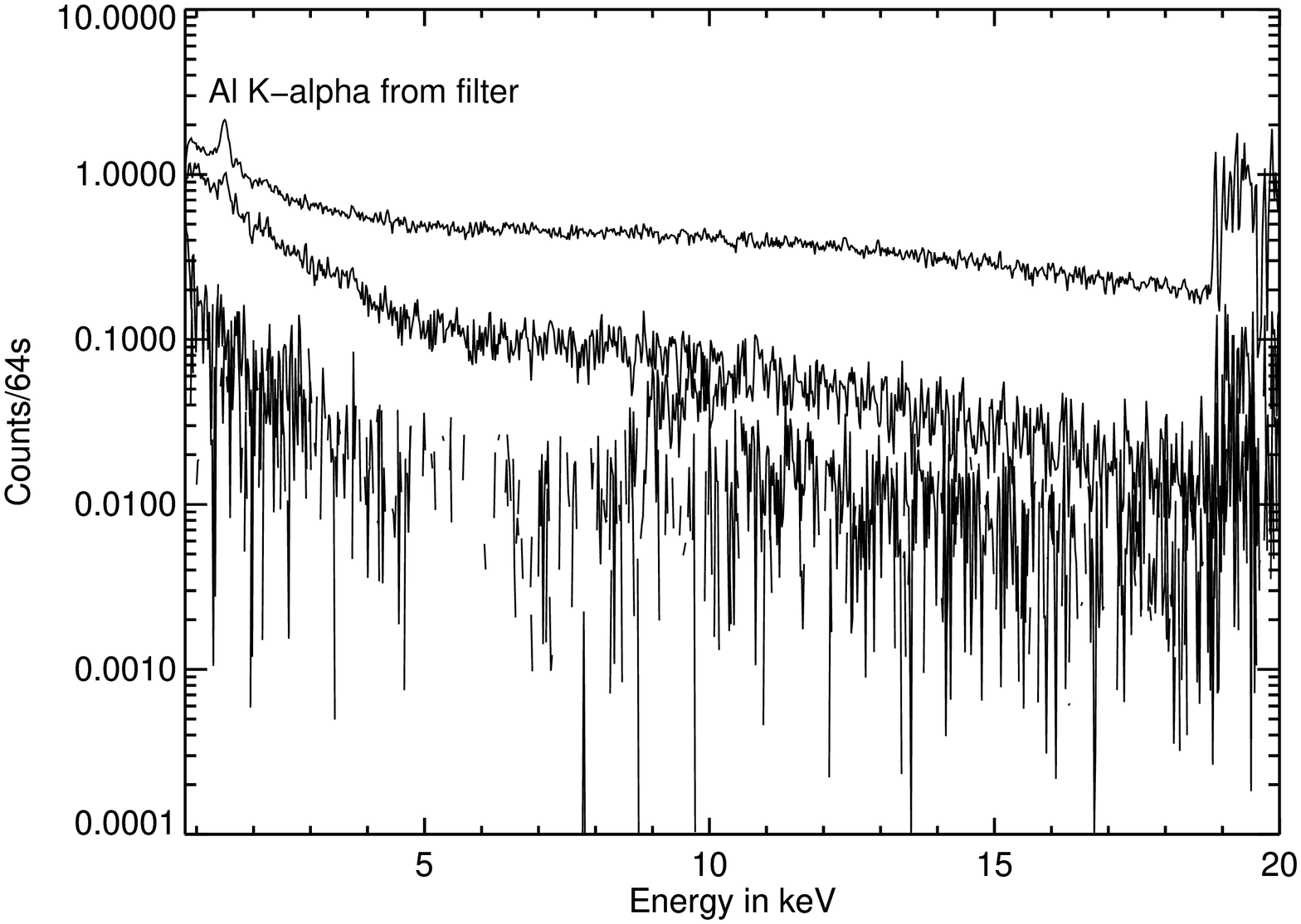}
\end{center}
\caption{Change in the background spectrum as C1XS passes through the geotail between 6 Feb 2009 and 10 Feb 2009. It can be seen that the spectrum gets harder (from bottom to the top) during the passage.}
\end{figure}
The instrument in orbit around the Moon encounters solar wind particles which interact in the detector material. These interactions form the background observed as a continuum throughout the observations. As the Moon goes through the geotail once in $\sim$29 days, the background spectrum changes considerably in shape as well as strength (Figure 4).  We find this increase during six of the total of nine geotail passes C1XS encountered (since the exact transit trajectory across the geotail cannot be determined, C1XS may not have always encountered the central core of the geotail where particle flux is expected to be even more enhanced.).
 A sudden increase in flux and presence of an enhanced hard x-ray spectral component ($>$a few keV) is observed when C1XS enters the geotail (approximately 6 days around full moon) and varies until the spacecraft leaves the region. From the whole mission data, we find that the background remains remarkably steady except for the geotail passes. 
\par The particles also excite characteristic x-rays from atoms it encounters including the aluminum filters (a maximum of $\sim$ 0.5 photons/cm$^2$/s is observed) placed in front of the detectors to block visible light. The Al x-rays from the filter are indistinguishable from Al XRF from the lunar surface. All the solar flares (the corresponding lunar data from which we infer composition) coincidentally occurred while C1XS was in the geotail. Thus modeling the detector background including derivation of contribution from Al x-ray photons from the filter proved challenging.
\par Al filters are still the best choice as far as the primary purpose of blocking visible light with minimal attenuation of the x-ray signal is concerned. Be filters as thin as 2 $\mu$m would be required to get the same performance as 0.2 $\mu$m Al deposited on a low Z substrate which is difficult to implement in practice (because of problems in procurement, handling, mounting etc). In CLASS, we will use a single Al filter of 0.2 $\mu$m as opposed to two in C1XS, to reduce contamination from the filter. We have also planned measurements of particle induced x-ray emission (PIXE) from the filter as part of ground calibration. Thus a quantitative estimate of Al x-rays from the filter for various particle spectra (through a combination of tests and simulations with the Monte Carlo toolkit GEANT4 \citep{Agostinelli03} ) would be used to aid lunar spectral analysis.

\subsection{Instrument calibration}
CLASS will undergo a detailed ground calibration similar to C1XS \citep{Naren2010} with emphasis on deriving the following

\begin{itemize}
\item Absolute quantum efficiency from 0.9-10 keV.
\item Spectral re-distribution function (SRF) from 0.9-10 keV as a function of temperature and count rate.
\item Test SCD for light leak effects due to reduced Al foil thickness.
\item Estimate PIXE contribution from Al filter from ground calibration.
\item An analytical code for simulation of the detector response has been developed \citep{Ath2012} and will be used for characterizing the response of the detectors along with laboratory measurements.

\end{itemize}

\par Onboard calibration will be done with the Fe-55 radioisotopes on the instrument door once during commissioning. We find from C1XS data that closed door condition produced calibration at lower temperature compared to nominal temperature during science observation. This could arise from the absence of lunar reflected heat load on the detector when the door is closed.
 Hence, after the door opens, calibration will use lunar XRF lines.

\subsection{Solar spectra}

An X-ray Solar Monitor is being developed at the Physical Research Laboratory, India \citep{sh2012} for measuring simultaneous solar spectra in the 1-15 keV. Solar spectral analysis will be carried out with detailed models in CHIANTI \citep{Dere09} as was done for the similar instrument in Chandrayaan-1 \citep{Nrrev}. 

\subsection{XRF intensity: Dependencies}

X-ray fluorescence intensity depends on a variety of factors other than solar x-ray spectra such as particle size, geometry of observation and the mineral matrix \citep{Mar2008, nr2008, wdr2011}. It is very difficult to deconvolve the absolute flux from lunar elements uniquely given the multi parameter dependencies. Therefore often in experiments for elemental mapping, the measured counts in the detector are calibrated against ground truths and abundances derived (for example \citep{yamashita2012}). While this simplifies the problem, it assumes that the range in abundances are confined to values observed in returned samples. Given the fact that these samples are from only nine locations on the Moon, this assumption  may not be always valid.
\par We plan to do   detailed investigations of dependencies of each of the parameters. Some of these correction were already implemented in the XRF code 'x2abundance' \citep{Ath2013} which will be further refined.
\par Efforts to derive correlations between spectral reflectance in IR and XRF are also in the pipeline which would provide a framework for such correlated analysis in future.   
\section{Particle induced X-ray emission from lunar surface}
Particles in the lunar environment can excite characteristic x-rays from the surface. Such observations can enable lunar elemental mapping  observations even during the night side of the orbit. Recently such measurements by the x-ray spectrometer on MESSENGER were used to derive surface abundances \citep{Starr2012} of Mercury. C1XS observed at least one such event where x-ray lines from Mg, Al and Si induced by particles were observed. This observation occurred during the dayside but solar flux levels  were very low to induce XRF. Kaguya-XRS also has observed PIXE from the onboard calibration plate \citep{okada10}. However, the C1XS PIXE spectrum from the lunar surface  is the first of its kind. Complications in the analysis of such spectra such as unknown incident particle spectrum, response of the detector to particles etc prevented any further analysis to derive elemental composition. 
\par A particle spectrometer alongside an XRF experiment would be a useful complementary instrument for utilizing PIXE data from lunar surface. Both XRF and PIXE photons from the Moon can be used to provide elemental maps enhancing coverage. In CLASS, studies are underway to use one of the swept charge devices coupled with a scintillator as a particle detector. The scintillator converts the energy deposited into an 'extended' light source on the SCD and currently efforts are underway to optimise the particle detection efficiency. 
\begin{figure}
\begin{center}
\includegraphics[height=12cm,width=10cm,angle=-90]{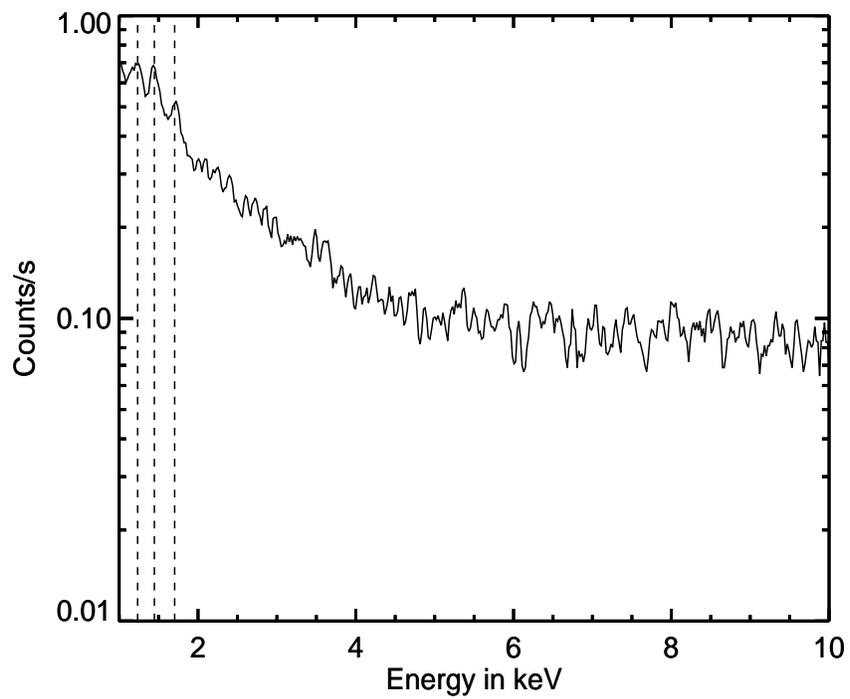}
\end{center}
\caption{Particle induced X-ray emission from Mg, Al and Si from lunar surface observed in C1XS (11 January 2009 01:45:01 to 01:59:03  UTC)}
\end{figure}
\section{Summary}

CLASS is expected to provide global maps of major elements from Na to Fe at resolutions of a few tens of kilometers. Together with mineralogical data this would provide a comprehensive picture of lunar surface chemistry.  We have used results from C1XS experiment flown on Chandrayaan-1 to refine the design as well as analysis methods for CLASS. We report on the particle induced x-ray emission from lunar surface observed with C1XS and emphasize on the possibility of combined particle measurements in order to use PIXE data along with solar induced XRF. CLASS will have a particle detector as a complimentary experiment in order to achieve the goal of mapping lunar chemistry using PIXE.

\section{Acknowledgements}
We thank Prof. Ian Crawford at Birkbeck College, London and the anonymous referee for their comments which has greatly improved the paper.






\begin{thebibliography}{}

\bibitem[Agostinelli et al.(2003)]{Agostinelli03}
Agostinelli, S., Allison, J., Amako, K., et al., Geant4: a Simulation Toolkit, NIM A, 506, 250-303, 2003.

\bibitem[Athiray et al.(2012)]{Ath2012}
Athiray, P.S., Narendranath, S., Sreekumar, P., Gow, J.,  Radhakrishna, V., Babu, B.R.S., Modeling charge transport in swept charge devices for X-ray spectroscopy ,Proc. of SPIE Vol.8453, 84532L1-9, 2012.

\bibitem[Athiray et al.(2013)]{Ath2013}
 Athiray, P.S., Narendranath, S., Sreekumar, P., Dash, S., K., Babu, B.R.S., Validation of methodology to derive elemental abundances from X-ray observations on Chandrayaan-1, Planet. Space Sci, 75, 188-194, 2013.

\bibitem[Crawford et al.(2009)]{crawford2009}

Crawford, I.A., Joy, K.H., Kellett, B.J., Grande, M., Bhandari, N., Cook, A.C., L d'Uston, Fernandes, V.A., Gasnault, O., Goswami, J., Howe, C.J., Huovelin, J., Koschny, D., Lawrence, D.J., Maddison, B.J., Maurice, S., Narendranath, S., Pieters, C.M., Okada, T., Rothery, D.A., Russell, S.S., Sreekumar, P., Swinyard, B., Wieczorek, M., Wilding, M., The scientific rationale for the C1XS X-ray Spectrometer
on India's Chandrayaan-1 mission to the Moon, Planet. Space Sci, 57, 725-734, 2009.


\bibitem[Dere et al.(2007)]{Dere09}
Dere, K.P., Landi, E., Mason, H.E., Monsignori Fossi, B.C., Young, P.R., CHIANTI-
An atomic database for emission lines, A$\&$A, 129, 149-173, 1997.

\bibitem[Grande et al.(2009)]{Grande09}

Grande, M., Maddison, B.J., Howe, C.J., Kellett, B.J., Sreekumar, P., Huovelin, J., Crawford, I.A., L d'Uston, Smith, D., Anand, M., Bhandari, N.,  Cook, A., Erd, C., Fernandes, V., Foing, B., Gasnault, O., Goswami, J.N., Holland, A., Joy, K.H., Koschny, D., Lawrence, D.J. , Maurice, S., Narendranath, S., Okada, T., Pieters, C.M., Rothery, D., Russell, S.S., Shrivastava, A., Swinyard, B.,  Wieczorek, M., Wilding, M., The C1XS X-ray Spectrometer on Chandrayaan-1, Planet. Space Sci, 57, 717-724, 2009.

\bibitem[Greenhagen et al.(2010)]{Green2010}
Greenhagen, B.T., Lucey, P.G., Wyatt, M., Glotch, T., Allen, C., Arnold, J., Bandfield, L., Bowles, N.E., Hanna, D., Hayne, P., Song, E., Thomas, I.R., Paige, D.A., Global silicate mineralogy of the Moon from the diviner lunar radiometer, Science, 329, 1507-1509, 2010.


\bibitem[Holland (2007)]{Br}
Holland, A., HXMT CCD236 SCD : Initial Test Report, 2007.


\bibitem[Lowe et al.(2001)]{L2001}
Lowe, B.G., Holland, A.D., Hutchinson, I.B., Burt, D.J., Pool, P.J., Swept Charge Devices, NIM A, 458, 568-579, 2001.


\bibitem[Maruyama et al.(2008)]{Mar2008}
Maruyama, Y., Ogawa, K., Okada, T., Kato, M., Laboratory experiments of particle size effect in X-ray fluorescence and implications to remote X-ray spectrometry of lunar regolith  surface, Earth Planets Space, 60, 293-297, 2008.

\bibitem[Naranen et al.(2008)]{nr2008}
Naranen, J., Parviainen, H., Muinonen, K., Carpenter, J., Nygård, K., Peura, M., Laboratory studies into the effect of regolith on planetary X-ray fluorescence spectroscopy, Icarus 198, 408-419, 2008.



\bibitem[Narendranath et al.(2010)]{Naren2010}

Narendranath, S., Sreekumar, P., Maddison, B.J., Howe, C.J., Kellett, B.J., Wallner, M., Erd, C., 
Weider, S.Z., Calibration of the C1XS instrument on Chandrayaan-1, NIM A , 621, 344-353, 2010.

\bibitem[Narendranath et al.(2011)]{Naren2011}
Narendranath, S., Athiray, P.S., Sreekumar, P., Kellett, B.J., Alha, L., Howe, C.J., Joy, K.H.,  Grande, M.,
Huovelin, J., Crawford, I.A., Unnikrishnan, U., Lalita, S., Subramaniam, S.,  Weider, S.Z., Nittler, L.R.,
Gasnault, O., Rothery, D., Fernandes, V.A., Bhandari, N., Goswami, J., N., Wieczorek, M., and the C1XS team,
Lunar X-ray fluorescence observations by the Chandrayaan-1 X-ray Spectrometer (C1XS): Results
from the nearside southern highlands, Icarus,  214, 53-66, 2011.

\bibitem[Narendranath et al.(2013)]{Nrrev}

Narendranath, S., Sreekumar, P., Alha, L., Sankarasubramanian, K., Huovelin, J., Athiray, P.S., Elemental abundances in the solar corona as measured by the X-ray Solar Monitor on Chandrayaan-1, Solar Physics, under review, 2013.








\bibitem[Narendranath et al.(2011)]{Naren2011}
Narendranath, S., Athiray, P.S., Sreekumar, P., Kellett, B.J., Alha, L., Howe, C.J., Joy, K.H.,  Grande, M.,
Huovelin, J., Crawford, I.A., Unnikrishnan, U., Lalita, S., Subramaniam, S.,  Weider, S.Z., Nittler, L.R.,
Gasnault, O., Rothery, D., Fernandes, V.A., Bhandari, N., Goswami, J.N., Wieczorek, M.A., and the C1XS team,
Lunar X-ray fluorescence observations by the Chandrayaan-1 X-ray Spectrometer (C1XS): Results
from the nearside southern highlands, Icarus,  214, 53-66, 2011.

\bibitem[Ohtake et al.(2009)]{Ohtake2009}
Ohtake, M., Matsunaga, T., Haruyama, J., et al., The global distribution of pure anorthosite on the Moon,
Nature, 461, 236-240, 2009.

\bibitem[Okada et al.(2009)]{okada10}
Okada, T., Shirai, K., Yamamoto, Y., Arai, T., Ogawa, K., Shiraishi, H., Iwasaki, M., Kawamura, T., Morito, H., Grande, M., Kato, M., X-ray fluorescence spectrometry of Lunar surface by XRS onboard SELENE (Kaguya), Trans.JSASS, 7, 39-42, 2009.

\bibitem[Pieters et al.(2009)]{pieters2009}
Pieters, C.M., Goswami, J.N., Clark, R.N., Annadurai, M., Boardman, J., Buratti, B., Combe, J.P., Dyar, M.D., Green, R., Head, J.W., Hibbitts, C., Hicks, M., Isaacson, P., Kilma, R., Kramer, G., Kumar, S., Livo, E., Lundeen, S., Malaret, E., McCord, T.,
Mustard, J., Nettles, J., Petro, N., Runyon, C., Staid, M., Sunshine, J., Taylor, L.A.,
Tompkins, S., and Varanasi, P., Character and Spatial Distribution of OH/H2O on the Surface of the Moon Seen by M3 on Chandrayaan I, Science,
287, 568-572, 2009.

\bibitem[Pieters et al.(2011)]{pieters2011}
Pieters, C.M., Besse, S., Boardman, J., Buratti, B., Cheek, L., Clark, R.N., Combe, J.P., Dhingra, D., Goswami, J.N., Green, R.O., Head, J.W., Isaacson, P., Klima, R., Kramer, G., Lundeen, S., Malaret, E., McCord, T., Mustard, T., Nettles, J., Petro, N., Runyon, C., Staid, M., Sunshine. J., Taylor, L.A.,Thaisen, K., Tompkins, S., Whitten, J., Mg-spinel lithology: A new rock type on the lunar farside, JGR, 116, E00G08, 2011. 

\bibitem[Radhakrishna V. et al.(2011)]{RK11}
Radhakrishna, V., Narendranath, S., Tyagi, A., Bug, M., Unnikrishnan, U., Kulkarni, R., Sreekantha, C.V., Kumar, G. Balaji, Athiray, P.S., Sudhakar, M., Manoj, R., Chetty, S.V., Thyagaraj, M.R., Howe, C.J., Gow, J., Sreekumar, P., The Chandrayaan-2 Large Area Soft X-ray Spectrometer (CLASS), LPSC abstract, 1708, 2011.


\bibitem[Shanmugham et al.(2012)]{sh2012}
Shanmugam, M., Vadawale, S., Acharya, Y.B., Goyal, S.K., Patel, A., Bhumi Shah and Murty, S.V.S., Solar X-ray monitor (XSM) onboard Chandrayaan-2 orbiter, LPSC abstract 1858, 2012.


\bibitem[Starr et al.(2012) ]{Starr2012}
Starr, R.D., Schriver, D., Nittler, L., Weider, S.Z., Byrne, P.K., Ho, G.C., Rhodes, E., Schlemm, C.E., Solomon, S.C., and Travnicek, P.M., Messenger detection of electron induced X-ray fluorescence from Mercury's surface, JGR, 117, E00L02, 2012.

\bibitem[Weider et al.(2011)]{wdr2011}

Weider, S.Z., Swinyard, B.M., Kellett, B.J., Howe, C.J., Joy, K.H., Crawford, I.A., Gow, J., Smith, D.R., 2011. Planetary X-ray fluorescence analogue laboratory experiments and an elemental abundance algorithm for C1XS, Planet. Space Sci, 59, 1393-1407, 2011. 

\bibitem[Weider et al.(2012)]{Wiederb12}
Weider, S.Z., Kellett, B.J., Swinyard, B.M., Crawford, I.A., Joy, K.H., Grande, M., Howe, C.J., Huovelin, J., Narendranath, S., Alha, L., Anand, M., Athiray, P.S., Bhandari, N., Carter, J.A., Cook, A.C., D'Uston, L.C., Fernandes, V.A., Gasnault, O., Goswami, J.N., Gow, J.P.D., Holland, A.D., Koschny, D., Lawrence, D.J., Maddison, B.J., Maurice, S., McKay, D.J., Okada, T., Pieters, C., Rothery, D.A., Russell, S.S., Shrivastava, A., Smith, D.R., Wieczorek, M., The Chandrayaan-1 X-ray Spectrometer, first results, Planet. Space Sci, 60, 217-228, 2012.

\bibitem[Wieczorek et al.(2006)]{Wz2006}
Wieczorek, M.A., Joliff, B.L., Khan, A., et al.,  The Constitution and Structure of the Lunar Interior,
Rev. Min. and Geochem. 60, 221-364, 2006.


\bibitem[Yamamoto et al.(2012)]{Yamamoto2012}
Yamamoto, S., Nakamura, R., Matsunaga, T., Ogawa, Y., Ishihara, Y., Morota, T., Hirata, N., Ohtake, M.,  Hiroi, T., Yokota, Y., and Haruyama, J., Massive layer of pure anorthosite on the Moon, GRL, 39, L13201, 2012.

\bibitem[Yamashita et al.(2012)]{yamashita2012}
Yamashita, N., Gasnault, O., Forni, O.C. d'Uston, Reedy, R.C., Karouji, Y., Kobayashi, S., Hareyama, M., Nagaoka, H., Hasebe, N., Kim, K., J., The global distribution of calcium on the Moon: Implications for high-Ca pyroxene in the eastern mare region, Earth and Planetary Science Letters 353, 93-98, 2012.
\end{thebibliography}
\end{document}